\documentclass[letterpaper]{article} 
\usepackage{aaai2027}           
\usepackage[hyphens]{url}      
\usepackage{graphicx}           
\usepackage{natbib}             
\usepackage{caption}            
\usepackage{amsmath,amssymb,amsthm}
\usepackage{booktabs}

\newcommand{\Wtwo}{W_2}
\newcommand{\R}{\mathbb{R}}
\newcommand{\Dh}{\bar D_{\mathrm{half}}}
\newcommand{\Dn}{\bar D_{\mathrm{null}}}

\nocopyright

\title{Topological Neural Collapse}

\author{
    Nicholas Enbar-Salo\textsuperscript{\rm 1},
    Jeova Farias Sales Rocha Neto\textsuperscript{\rm 1}
}
\affiliations{
    \textsuperscript{\rm 1}Bowdoin College, Brunswick, ME, USA\\
    nenbarsalo@bowdoin.edu, j.farias@bowdoin.edu
}

\begin{document}
\maketitle

\begin{abstract}
Neural collapse (NC) predicts that, late in training, last-layer class
features concentrate at their means and the means form a simplex
equiangular tight frame (ETF). We ask whether NC also leaves a
signature in the functional connectome of each layer: the graph of
Pearson correlations between hidden-unit activations, built per class
and summarized by persistent graph homology. Using deep MLPs, we observe two topological
phenomena that co-occur with the well-studied geometric NC in the final
layers during the terminal phase. Contrary to NC1, the split-half
variability of a class connectome increases (TNC1). The mean pairwise
\(H_0\) distance \(\bar{D}\) between class connectomes also rises
(TNC2), but unevenly across class pairs, unlike the equidistance of
NC2. We find that 24--56\% of last-layer units become dead for some
classes yet stay active for others, which inflates \(\bar{D}\). After
correcting for these units, both signatures remain elevated. These
results from MNIST, Fashion-MNIST, CIFAR-10, and CIFAR-100 provide preliminary evidence that a topological mechanism may
accompany the geometric signatures of NC.
\end{abstract}

\section{Introduction}
Neural collapse \citep{papyan2020prevalence} describes the geometry of
last-layer features in the terminal phase of training (TPT): within-class
variability vanishes (NC1), class means form a simplex equiangular tight frame (NC2), the
classifier aligns with the means (NC3), and the classifier utilizes a
nearest-class-mean rule (NC4). The phenomenon matters because collapse
has been linked to generalization and the ability of a network to transfer to new classes
\citep{papyan2020prevalence,galanti2022role}. Subsequent work has also
found NC, though weaker, in intermediate layers
\citep{parker2023neural}.

Inspired by functional connectomes in functional magnetic resonance imaging (fMRI) studies of biological
networks, \citet{zhang2023functional} proposed examining a
network through its \emph{functional connectome}: a weighted graph of
Pearson correlations among unit activations, summarized by persistent homology. In this paper we construct class-specific, per-layer connectomes and ask whether NC leaves a topological trace. Here, we observe two
topological patterns that co-occur with NC1 and NC2:
\textbf{(TNC1)} each class connectome becomes more variable within its
own class, and
\textbf{(TNC2)} the class-connectome means move farther apart in the
terminal phase of training. 

\paragraph{Contributions.}
(i)~We define and observe the two connectome signatures and track them, together
with NC1--NC4, across layers and training checkpoints.
(ii)~We find that both TNC1 and TNC2 increase as training progresses, deviating from NC1 and violating the equidistance property of NC2 
(iii)~We report an unexpected by-product: a surge of class-selective
dead units in the terminal phase.

\begin{figure*}[t]
\centering
\includegraphics[width=0.85\textwidth]{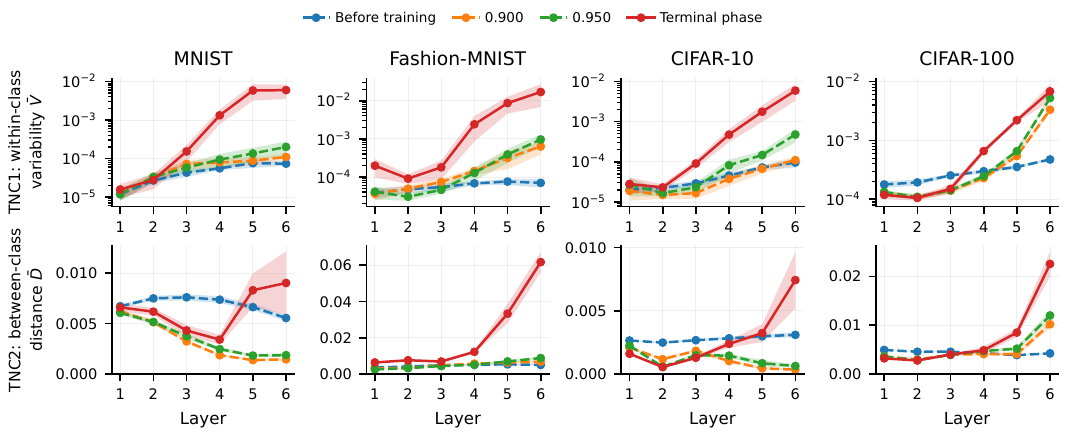}
\caption{Layer-wise connectome signatures (mean
$\pm$ s.d.). \emph{Top}: split-half
within-class variability $\bar{V}$ (TNC1), log scale \emph{Bottom}: mean pairwise $H_0$ distance
$\bar{D}$ between class connectomes (TNC2).}
\label{fig:sig}
\end{figure*}

\begin{table*}[t]
\centering
\small
\begin{tabular}{lcccccc}
\toprule
 & \multicolumn{2}{c}{Collapse} & \multicolumn{2}{c}{TNC1: within-class variability}
 & \multicolumn{2}{c}{TNC2: between-class distance} \\
\cmidrule(lr){2-3}\cmidrule(lr){4-5}\cmidrule(lr){6-7}
Dataset & NC1 & NC2 & $\bar{V}\,(\times 10^{-4})$ & $\bar{V}/\Dh$ & $\bar{D}$ & CV$^\ast$ \\
\midrule
MNIST         & 1.25 $\to$ .004 & .70 $\to$ .27  & 0.7 $\to$ 59  & .01 $\to$ .72 & .0056 $\to$ .0090 & 1.64 $\to$ 1.15 \\
Fashion-MNIST & 2.85 $\to$ .002 & .84 $\to$ .19  & 0.7 $\to$ 169 & .01 $\to$ .26 & .0051 $\to$ .0618 & 1.14 $\to$ .79  \\
CIFAR-10      & 21.3 $\to$ .034 & 1.01 $\to$ .19 & 1.0 $\to$ 59  & .03 $\to$ .63 & .0031 $\to$ .0074 & .98 $\to$ .82   \\
CIFAR-100     & 48.1 $\to$ .0013 & 1.27 $\to$ .19 & 4.7 $\to$ 68 & .11 $\to$ .28 & .0043 $\to$ .0225 & 1.01 $\to$ .96 \\
\bottomrule
\end{tabular}
\caption{Last hidden layer (fc6), mean over 10 seeds; two-value entries
are initialization $\to$ terminal phase (TPT). $^\ast$Coefficient of
variation of the pairwise distances.}
\label{tab:main}
\end{table*}

\section{Background: Persistent Graph Homology}
Let $X_c \in \R^{n_c \times p}$ be the activations of a layer with $p$
units on the $n_c$ training samples of class $c$. The class connectome
$G_c$ is the complete graph on the $p$ units with edge weights
$w_{ij}=|\mathrm{r}(X_{c,\cdot i},X_{c,\cdot j})|$, where r is the Pearson correlation.
Following \citet{songdechakraiwut2023topological}, let $G_\epsilon$
retain only edges whose weight exceeds $\epsilon$. As $\epsilon$ grows,
the number of connected components $\beta_0(G_\epsilon)$ increases.
The 0-dimensional persistence diagram consists of the $p-1$ weights of
the maximum spanning tree (MST), the \emph{births}, while the
1-dimensional diagram consists of the remaining non-tree weights (the
\emph{deaths}). The births represent the births of connected components, while the deaths represent the death of cycles.

\citet{songdechakraiwut2025functional}, and later adapted by \citet{wu2025dataefficient} to track intermediate stages of training, demonstrated that, for the connectomes
$G_c$ and $G_{c'}$ of two classes $c \neq c'$ with the same number of nodes,
the squared 2-Wasserstein distance ($W_2^2$) between their diagrams reduces to the
squared Euclidean distance between the sorted birth vectors $\mathbf{b}^{(c)}$
(for $H_0$) or sorted death vectors $\mathbf{d}^{(c)}$ (for $H_1$):
\begin{align}
\Wtwo^2\bigl(H_0^{(c)},H_0^{(c')}\bigr) &= \|\mathbf{b}^{(c)}-\mathbf{b}^{(c')}\|_2^2,\\
\Wtwo^2\bigl(H_1^{(c)},H_1^{(c')}\bigr) &= \|\mathbf{d}^{(c)}-\mathbf{d}^{(c')}\|_2^2.
\end{align}

In this paper, we report the per-entry mean squared distance
$\mathrm{MSD}=\Wtwo^2/m$ for TNC2, where $m$ is the length of the vector.
We restrict our analysis to $H_0$ and $H_1$ (connected components and cycles), following \citet{songdechakraiwut2025functional}, who use these two dimensions to characterize connectome topology. For simplicity, we mainly focus on $H_0$, while $H_1$
results are summarized in the Discussion. 

\section{Topological Connectome Signatures}
\paragraph{TNC1: within-class variability.}
To obtain a topological analogue of within-class variability (NC1) we use
split-half resampling, a method adopted from the fMRI literature to track measurement reliability \citep{teeuw2021reliability, walther2016reliability}. For each seed we randomly partition every class
into two halves and build a connectome from each half. Let $V_c$ be the
MSD between the two connectomes of class $c$, and let $\bar{V}$ be
the mean of $V_c$ over classes. We then compare $\bar{V}$ with $\Dh$, the mean MSD
between half-size connectomes of different classes. The ratio
$\bar{V}/\Dh$ is the topological counterpart of the NC1 ratio of
within- to between-class spread \cite{papyan2020prevalence}. To see whether the observed variability is due to noise alone, we also compute
$\Dn$, the mean MSD among $C$ random, class-agnostic groups of the
same size to determine the expected noise in the variability. Split-half variability is required because a connectome is a
property of a group of samples. Unlike the geometric features of NC1, TNC1 cannot be
measured on individual samples.

\paragraph{TNC2: between-class distance.}
$\bar{D}$ is the mean of $\mathrm{MSD}(H_0^{(c)},H_0^{(c')})$ over all
class pairs. If the classes are exchangeable (as expected under an
ETF in NC2) pairwise distances should be nearly equal. We quantify departures
from equality by the coefficient of variation (CV) of the pairwise
distances: the standard deviation divided by the mean. The CV equals
zero when every pair is equally far apart and grows as the distances
become more uneven. Our goal is that, by measuring the magnitude of pairwise distances as well as the variation in pairwise distances, we can observe a topological analogue to NC2.

\section{Experiments}
\paragraph{Setup.}
We train MLPs with six hidden layers of width 4096 (ReLU), following
\citet{parker2023neural}, using MSE loss. Optimization
uses SGD with momentum 0.9, a one-cycle schedule (peak learning rate
0.05), batch size 128, weight decay \(10^{-5}\), and 300 epochs. MSE is
preferred over cross-entropy because it both encourages NC and is mathematically clearer \citep{han2022neural}. We use 500
(CIFAR-100), 5000 (MNIST, CIFAR-10) or 6000 (Fashion-MNIST) training
images per class and average over 10 random seeds.

All quantities are evaluated on the training set at initialization, multiple intermediate stages of training, and at the
terminal phase. At the same checkpoints we compute
NC1 \(=\operatorname{tr}(\Sigma_W\Sigma_B^\dagger)/C\),
NC2 (Frobenius distance of the normalized Gram matrix of centered class
means to that of a simplex ETF), NC3 (distance between the normalized
classifier and class-mean matrices), and NC4 (disagreement with the
nearest-class-mean rule). See \citet{papyan2020prevalence} and
\citet{han2022neural} for further explanation on the derivations. Throughout these experiments, we name hidden layers 1 to 6 fc1 to fc6.

\paragraph{The networks collapse.}
At the final hidden layer (fc6), NC1 falls by two to three orders of magnitude and NC2 by
\(2.6\)--\(5.2\times\) (Table~\ref{tab:main}); NC4 falls from roughly
\(0.9\) to below \(10^{-3}\). After training begins, NC1 decreases both
with depth and with continued training at every layer.

\begin{figure}[t]
\centering
\includegraphics[width=0.9\columnwidth]{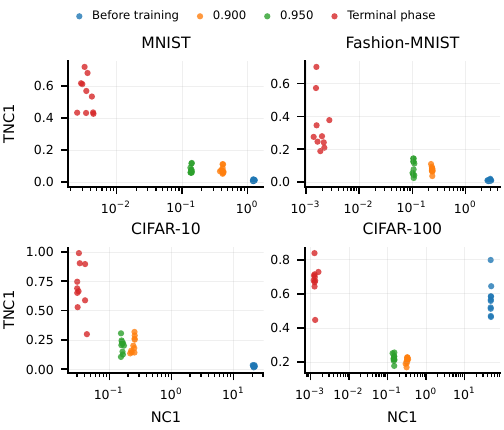}
\caption{NC1 against \(\bar{V}/\Dh\) (TNC1) at fc6. One point per seed and
checkpoint.}
\label{fig:nc}
\end{figure}

\paragraph{TNC1: within-class variability rises as NC1 falls.}
At fc6, \(\bar{V}\) grows by at most \(14\times\) before the terminal phase, and then reaches \(81\times\), \(242\times\), \(61\times\) and \(14\times\) its
initial value in the terminal phase (Figure~\ref{fig:sig}, top). The
increase becomes more pronounced with depth from fc3 onward (fc4 on CIFAR-100). The ratio
\(\bar{V}/\Dh\) rises from \(0.01\)--\(0.11\) to \(0.26\)--\(0.72\)
(Table~\ref{tab:main}) while NC1 falls. Plotting NC1 against
\(\bar{V}/\Dh\) yields a strong anti-correlation
(Spearman \(\rho=-0.90\), \(-0.89\), \(-0.89\), \(-0.21\); Figure~\ref{fig:nc}).
This is the opposite of the behavior expected from NC1, under which
within-class variability steadily declines.

\paragraph{TNC2: distance drops early, then rises in the terminal phase.}
In the intermediate stages of training, the fc6 value of \(\bar{D}\) falls \(3.8\times\)
on MNIST and \(8.6\times\) on CIFAR-10, while on Fashion-MNIST it had risen slightly
(Figure~\ref{fig:sig}, bottom; Table~\ref{tab:main}). In the terminal phase
\(\bar{D}\) rises on every dataset, finishing \(1.6\times\), \(12\times\)
and \(2.4\times\) above its initial value. On MNIST and CIFAR-10 the
rebound is concentrated in the last two layers: at fc1--fc4 the
terminal-phase distance remains below its initial value. On
Fashion-MNIST the terminal-phase distance exceeds the initial value at
every layer. The CV of the pairwise distances stays between \(0.7\) and
\(1.2\) after initialization, indicating that the pairwise connectome
distances never approach uniformity as expected under NC2. On CIFAR-100, \(\bar{D}\) shows no early drop: the fc6 value already
rises \(2.4\times\) during the intermediate training stages and finishes \(5.3\times\) above its
initial value, with the terminal-phase rebound confined to fc4--fc6. Its
CV stays between \(0.93\) and \(1.14\) at fc6, again far from uniform.

\section{Class-Selective Dead Units Confounding Class Connectome Distance}\label{sec:dead}

\begin{figure*}[t]
\centering
\includegraphics[width=0.79\textwidth]{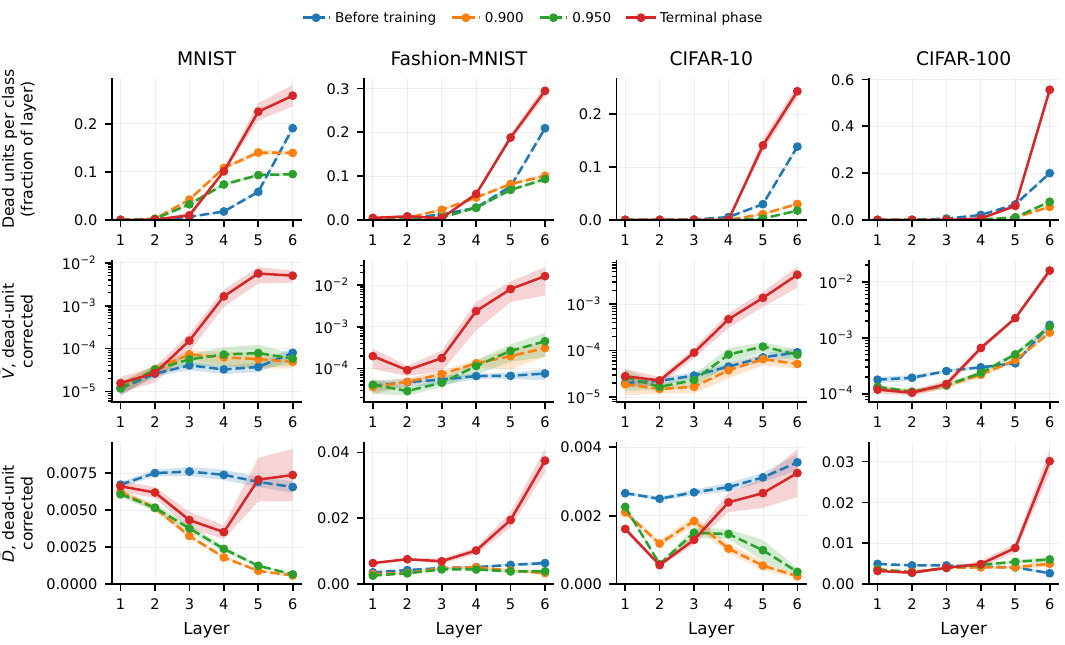}
\caption{Units dead for a class as a
fraction of the layer, and dead-unit corrected values for TNC1 and TNC2, (mean \(\pm\) s.d.).}
\label{fig:dead}
\end{figure*}

\begin{table*}[t]
\centering
\small
\begin{tabular}{lccccccc}
\toprule
 & \multicolumn{2}{c}{Dead units, fc6} & \multicolumn{2}{c}{$\bar{D}$, all units}
 & \multicolumn{3}{c}{$\bar{D}$ on the corrected unit set, fc6} \\
\cmidrule(lr){2-3}\cmidrule(lr){4-5}\cmidrule(lr){6-8}
Dataset & per class: init $\to$ 90\% $\to$ TPT & selective$^\ast$ & init $\to$ TPT & $\rho$ & init $\to$ TPT & $\rho$ & units$^\dagger$ \\
\midrule
MNIST         & 781 $\to$ 570 $\to$ 1058 & 2446 & .0056 $\to$ .0090 & .97 & .0066 $\to$ .0074 & .95 & 2577 \\
Fashion-MNIST & 857 $\to$ 412 $\to$ 1207 & 3230 & .0051 $\to$ .0618 & .99 & .0064 $\to$ .0375 & .98 & 2232 \\
CIFAR-10      & 569 $\to$ 124 $\to$ 998  & 2082 & .0031 $\to$ .0074 & .94 & .0036 $\to$ .0033 & .57 & 2624 \\
CIFAR-100     & 819 $\to$ 227 $\to$ 2278 & 4095 & .0043 $\to$ .0225 & .95 & .0027 $\to$ .0301 & .40 & 161 \\
\bottomrule
\end{tabular}
\caption{Dead units and the dead-unit correction at fc6, mean over 10
seeds. Dead units are counts per class out of 4096. $^\ast$Units dead
in some but not all classes at TPT. $\rho$: Spearman correlation over
class pairs between the dead-unit gap and the $H_0$ distance at TPT.
$^\dagger$Size of the corrected unit set.}
\label{tab:corr}
\end{table*}

Although between-class distances initially decrease, they balloon in the
terminal phase. While investigating the cause of this increase, we noticed a
substantial population of dead neurons in fc6 during the terminal
phase (units that never activate for any sample; Figure~\ref{fig:dead}). Such
units have zero variance, so their correlations are undefined. For this study, we set those correlations to zero. We call a unit \emph{dead in class \(c\)}
and write \(k_c\) for their number. Prior work has attributed this artifact
to ReLU, which collapses some units to zero variance
\citep{lu2020dying}.

\paragraph{Dead units surge in the terminal phase and are class-selective.}
At fc6, the dead fraction 
then jumps in the terminal phase (Figure~\ref{fig:dead}):
\(19\to26\%\) on MNIST, \(21\to29\%\) on Fashion-MNIST,
\(14\to24\%\) on CIFAR-10, and \(20\to56\%\) on CIFAR-100. Most of these units are not dead across
every class: 51--79\% of the layer is dead for some classes and alive
for others. Thus, the dead units are class-selective.

\paragraph{Dead units inflate the distance.}
A class with more dead units possesses more zero births and is therefore
farther from other classes regardless of the behavior of the remaining
units. Empirically, the Spearman correlation (over class pairs) between
the dead-unit gap \(|k_c-k_{c'}|\) and the fc6 distance rises from
\(0.12\)--\(0.65\) at initialization to \(0.94\)--\(0.99\) in the
terminal phase. Consequently, the dead units inflate between-class distances and confound any purely topological signature of NC
(Figure~\ref{fig:sig}).

\paragraph{Dead units are not the whole explanation.}
Here we recompute both signatures after retaining
only those neurons that are dead for fewer than 20\% of classes. After
this correction, the topological patterns persisted, though they were
attenuated. The fc6 distance still falls throughout the intermediate training stages by \(11\times\) on MNIST, \(1.9\times\) on Fashion-MNIST and
\(15\times\) on CIFAR-10 (Figure~\ref{fig:dead};
Table~\ref{tab:corr}). In the terminal phase it climbs back to its
initial value on MNIST and CIFAR-10 and to \(5.9\times\) that value on
Fashion-MNIST. Applying the 20\% rule independently at each checkpoint
yields the same pattern (except that Fashion-MNIST then shows no early
drop). Within-class variability on the corrected set of neurons rises to \(63\times\),
\(217\times\) and \(47\times\) its initial value, and
\(\bar{V}/\Dh\) rises from \(0.01\)--\(0.03\) to \(0.34\)--\(0.70\);
NC1 and \(\bar{V}/\Dh\) remain anti-correlated
(\(\rho=-0.86\), \(-0.88\), \(-0.82\)).
On CIFAR-100, the correction retains only 161 of 4096 units, so its
results are less reliable. On this set, \(\bar{D}\) shows no early drop
(rising \(1.9\times\) before terminal phase) and ends \(11\times\) above its
initial value; \(\bar{V}\) rises \(9.4\times\), but \(\bar{V}/\Dh\) moves
only from \(0.57\) to \(0.68\) and its anti-correlation with NC1 is weak
(\(\rho=-0.21\)).
Taken together, these results indicate that a mechanism beyond dead
units is driving the terminal-phase increase of between-class distances
in the final layers.
\section{Discussion and Limitations}
The networks collapse by the standard geometric metrics, yet in the
terminal phase both between-class distance and within-class variability
rise in the topological connectomes. Moreover, the increase in between-class
distance is not uniform across pairs, contrary to the prediction of NC2.
Results for \(H_1\) are similar: the fc6 \(H_1\) distance
ends \(1.9\)--\(6.6\times\) above its initial value, and the \(H_1\)
split-half ratio rises from at most \(0.04\) to \(0.20\)--\(0.94\) (Appendix~\ref{app:h1}).

We see two possible explanations. First, NC itself may be responsible,
but it may affect topology differently from geometry, simultaneously
increasing both between-class distance and within-class variability.
Second, it is possible that, when features collapse onto
non-negative class means, ReLU units can become silent for entire
classes. ReLU feature models predict collapsed means with little overlap
\citep{tirer2022extended,dang2024neural}, which would generate
class-selective dead units. Dying-ReLU effects already present at
initialization \citep{lu2020dying} may account for the dead units
observed before training begins. Future work should extend these results to other activation functions to see whether these results are due to ReLU effects or a consequence of NC.

\paragraph{Limitations.}
TNC1 relies on a single random split per seed. Future work should average
over multiple splits. All experiments use MSE loss and ReLU MLPs. We
did not vary sample size, activation function, or weight decay, any of
which could shed light on the mechanisms underlying these findings. Finally, the results are purely correlational, so no causal claim that the observed topological changes are produced by NC can be made.

\paragraph{Conclusion and Future Directions.} These results suggest that a
topological counterpart of NC may occur alongside geometric NC. Future work
should test whether they generalize to other loss functions and activation
functions, especially ones such as LeakyReLU and GELU that never output exact
zeros and so avoid the dead-unit confound. Inducing topological collapse
directly could also be tested for the generalization benefits associated
with NC.

\paragraph{AI Disclosure.}
We used Claude Opus 5.5 to edit and polish the grammar and clarity of the author-written text in this paper. All outputs were reviewed and edited by the authors, who take full responsibility for the content. Additionally, we used Claude Code for debugging. The authors validated and executed all final code used in our experiments, and all experimental design decisions were made by the authors.

\bibliography{references}

\appendix
\setcounter{secnumdepth}{2} 

\section{\(H_1\) Results}\label{app:h1}
This appendix repeats the \(H_0\) analysis of the main text with the \(H_1\)
death vectors \(\mathbf{d}^{(c)}\) in place of the birth vectors:
\(\bar{V}_1\), \(\bar{D}_{1,\mathrm{half}}\) and \(\bar{D}_1\) are defined as
\(\bar{V}\), \(\Dh\) and \(\bar{D}\), with the MSD taken over deaths.

\subsection{Connectome Signatures}\label{app:h1_sig}
\begin{figure*}[t]
\centering
\includegraphics[width=0.85\textwidth]{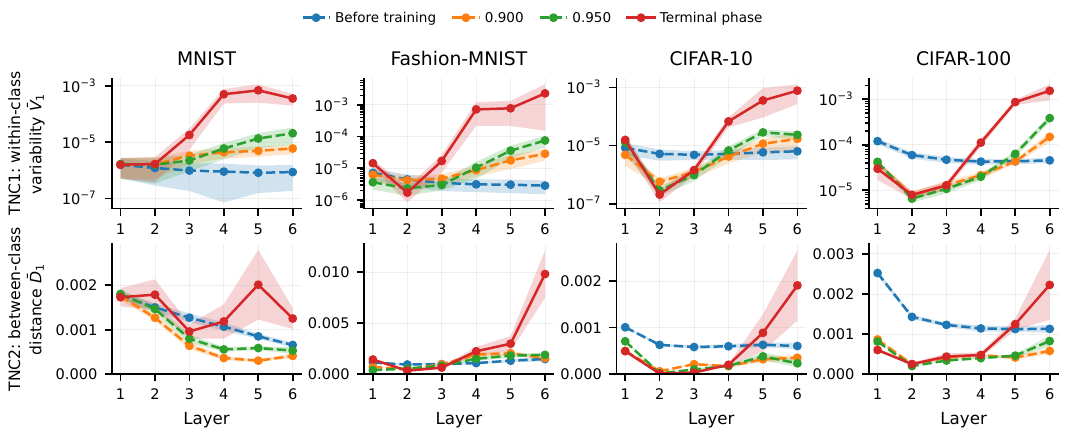}
\caption{\(H_1\) connectome signatures (mean \(\pm\) s.d.). \emph{Top}:
split-half within-class variability \(\bar{V}_1\) (TNC1), log scale.
\emph{Bottom}: mean pairwise \(H_1\) distance \(\bar{D}_1\) between class
connectomes (TNC2). Compare Figure~\ref{fig:sig}.}
\label{fig:sig_h1}
\end{figure*}

Figure~\ref{fig:sig_h1} and Table~\ref{tab:main_h1} are the \(H_1\)
counterparts of Figure~\ref{fig:sig} and Table~\ref{tab:main}. As for
\(H_0\), both signatures rise in the terminal phase at fc6: \(\bar{V}_1\)
grows \(34\)--\(805\times\) its initial value and \(\bar{D}_1\)
\(1.9\)--\(6.6\times\), while \(\bar{V}_1/\bar{D}_{1,\mathrm{half}}\) rises
from at most \(0.04\) to \(0.20\)--\(0.94\). At TPT the CV of the pairwise \(H_1\)
distances is \(1.0\)--\(1.9\), so the \(H_1\) distances are
no closer to uniform than the \(H_0\) distances.

\subsection{Dead-Unit Correction}\label{app:h1_fixed}
\begin{figure*}[t]
\centering
\includegraphics[width=0.85\textwidth]{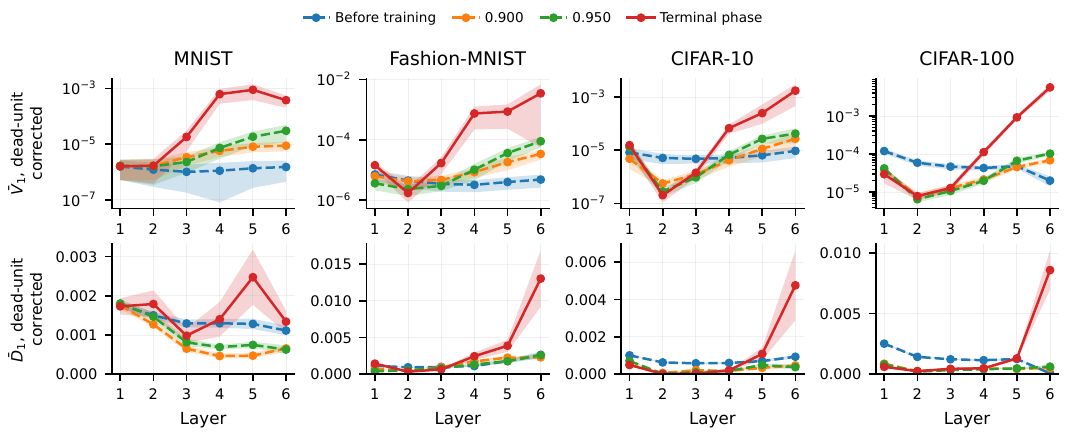}
\caption{\(H_1\) signatures after the dead-unit correction (mean \(\pm\)
s.d.). Compare the bottom two rows of Figure~\ref{fig:dead}.}
\label{fig:fixed_h1}
\end{figure*}

Figure~\ref{fig:fixed_h1} and Table~\ref{tab:corr_h1} repeat the dead-unit
correction of Section~\ref{sec:dead} on the same corrected unit sets. The dead-unit gap is
only weakly tied to the \(H_1\) distance (\(\rho=0.11\)--\(0.71\) at TPT,
against \(0.94\)--\(0.99\) for \(H_0\)), consistent with dead units mainly
adding zero births rather than changing the cycle structure. After the
correction, \(\bar{D}_1\) still ends above its initial value on every dataset
and \(\bar{V}_1/\bar{D}_{1,\mathrm{half}}\) still rises.

\begin{table*}[t]
\centering
\small
\begin{tabular}{lcccc}
\toprule
 & \multicolumn{2}{c}{TNC1: within-class variability}
 & \multicolumn{2}{c}{TNC2: between-class distance} \\
\cmidrule(lr){2-3}\cmidrule(lr){4-5}
Dataset & $\bar{V}_1\,(\times 10^{-6})$ & $\bar{V}_1/\bar{D}_{1,\mathrm{half}}$ & $\bar{D}_1$ & CV$^\ast$ \\
\midrule
MNIST         & 0.9 $\to$ 354  & .001 $\to$ .33 & .0007 $\to$ .0013 & 1.74 $\to$ 1.19 \\
Fashion-MNIST & 2.8 $\to$ 2291 & .002 $\to$ .20 & .0015 $\to$ .0098 & 1.17 $\to$ 1.02 \\
CIFAR-10      & 6.4 $\to$ 772  & .011 $\to$ .31 & .0006 $\to$ .0019 & 1.07 $\to$ 1.08 \\
CIFAR-100     & 45.5 $\to$ 1539 & .043 $\to$ .94 & .0011 $\to$ .0022 & 1.56 $\to$ 1.91 \\
\bottomrule
\end{tabular}
\caption{\(H_1\) counterpart of Table~\ref{tab:main}: last hidden layer
(fc6), all units, mean over 10 seeds; entries are initialization \(\to\)
terminal phase (TPT). $^\ast$Coefficient of variation of the pairwise
\(H_1\) distances.}
\label{tab:main_h1}
\end{table*}

\begin{table*}[t]
\centering
\small
\begin{tabular}{lcccccc}
\toprule
 & \multicolumn{2}{c}{$\bar{D}_1$, all units}
 & \multicolumn{4}{c}{Corrected unit set, fc6} \\
\cmidrule(lr){2-3}\cmidrule(lr){4-7}
Dataset & init $\to$ TPT & $\rho$ & $\bar{D}_1$: init $\to$ TPT & $\rho$
 & $\bar{V}_1\,(\times 10^{-6})$ & $\bar{V}_1/\bar{D}_{1,\mathrm{half}}$ \\
\midrule
MNIST         & .0007 $\to$ .0013 & .45 & .0011 $\to$ .0013 & .28 & 1.5 $\to$ 383  & .001 $\to$ .29 \\
Fashion-MNIST & .0015 $\to$ .0098 & .71 & .0026 $\to$ .0130 & .48 & 4.8 $\to$ 3507 & .002 $\to$ .22 \\
CIFAR-10      & .0006 $\to$ .0019 & .22 & .0009 $\to$ .0048 & .08 & 9.5 $\to$ 1734 & .010 $\to$ .29 \\
CIFAR-100     & .0011 $\to$ .0022 & .11 & .0001 $\to$ .0086 & .31 & 20.1 $\to$ 5602 & .307 $\to$ .79 \\
\bottomrule
\end{tabular}
\caption{\(H_1\) counterpart of Table~\ref{tab:corr}: the dead-unit
correction at fc6, mean over 10 seeds. $\rho$: Spearman correlation over
class pairs between the dead-unit gap and the \(H_1\) distance at TPT. The
corrected unit sets are those of Table~\ref{tab:corr}.}
\label{tab:corr_h1}
\end{table*}

\end{document}